\documentclass[aps,prl,preprint,nopacs,superscriptaddress]{revtex4}

\usepackage{graphicx}
\usepackage{verbatim}
\usepackage{mathrsfs}
\usepackage{gensymb}
\usepackage{color}
\usepackage{ulem}

\usepackage{amsmath,amsfonts,amssymb}
\usepackage{graphicx}

\newcommand{\bs}[1]{{\boldsymbol{#1}}}

\newcommand{\bee}{\begin{equation}}
\newcommand{\ee}{\end{equation}}
\def\3{2.8in}    
\def\2{2.5in}
\def\4{3.0in}\def \beq {\begin{equation}}
\def \eeq {\end{equation}}
\begin{document}

\title{Topological Dirac Surface States and Superconducting Paring Correlations in PbTaSe$_2$}

\author{Tay-Rong~Chang\footnote{These authors contributed equally to this work.}}
\affiliation {Department of Physics, National Tsing Hua University, Hsinchu 30013, Taiwan}
\affiliation { Laboratory for Topological Quantum Matter and Spectroscopy (B7), Department of Physics,  Princeton University, Princeton, New Jersey 08544, USA}

\author{Peng-Jen~Chen$^*$}
\affiliation{Department of Physics, National Taiwan University, Taipei 10617, Taiwan}
\affiliation{Nano Science and Technology Program, Taiwan International Graduate Program, Academia Sinica, Taipei 11529, Taiwan and National Taiwan University, Taipei 10617, Taiwan}
\affiliation{Institute of Physics, Academia Sinica, Taipei 11529, Taiwan}

\author{Guang~Bian$^*$}
\affiliation { Laboratory for Topological Quantum Matter and Spectroscopy (B7), Department of Physics,  Princeton University, Princeton, New Jersey 08544, USA}

\author{Shin-Ming Huang$^*$}
\affiliation {Centre for Advanced 2D Materials and Graphene Research Centre National University of Singapore, 6 Science Drive 2, Singapore 117546}
\affiliation {Department of Physics, National University of Singapore, 2 Science Drive 3, Singapore 117542}

\author{Hao~Zheng}
\affiliation { Laboratory for Topological Quantum Matter and Spectroscopy (B7), Department of Physics,  Princeton University, Princeton, New Jersey 08544, USA}

\author{Titus Neupert}
\address{
 Princeton Center for Theoretical Science, Princeton University, Princeton, New Jersey 08544, USA}

\author{Raman~Sankar}
\affiliation {Center for Condensed Matter Sciences, National Taiwan University, Taipei 10617, Taiwan}

\author{Su-Yang~Xu}
\affiliation { Laboratory for Topological Quantum Matter and Spectroscopy (B7), Department of Physics,  Princeton University, Princeton, New Jersey 08544, USA}

\author{Ilya~Belopolski}
\affiliation { Laboratory for Topological Quantum Matter and Spectroscopy (B7), Department of Physics,  Princeton University, Princeton, New Jersey 08544, USA}

\author{Guoqing Chang}
\affiliation {Centre for Advanced 2D Materials and Graphene Research Centre National University of Singapore, 6 Science Drive 2, Singapore 117546}
\affiliation {Department of Physics, National University of Singapore, 2 Science Drive 3, Singapore 117542}

\author{BaoKai~Wang}
\affiliation {Centre for Advanced 2D Materials and Graphene Research Centre National University of Singapore, 6 Science Drive 2, Singapore 117546}
\affiliation {Department of Physics, National University of Singapore, 2 Science Drive 3, Singapore 117542}
\affiliation {Department of Physics, Northeastern University, Boston, Massachusetts 02115, USA}

\author{Fangcheng~Chou}
\affiliation {Center for Condensed Matter Sciences, National Taiwan University, Taipei 10617, Taiwan}

\author{Arun~Bansil}
\affiliation {Department of Physics, Northeastern University, Boston, Massachusetts 02115, USA} 

\author{Horng-Tay~Jeng\footnote{jeng@phys.nthu.edu.tw}}
\affiliation {Department of Physics, National Tsing Hua University, Hsinchu 30013, Taiwan}
 \affiliation {Institute of Physics, Academia Sinica, Taipei 11529, Taiwan}
 
\author{Hsin~Lin\footnote{nilnish@gmail.com}}
\affiliation {Centre for Advanced 2D Materials and Graphene Research Centre National University of Singapore, 6 Science Drive 2, Singapore 117546}
\affiliation {Department of Physics, National University of Singapore, 2 Science Drive 3, Singapore 117542}

\author{M. Zahid~Hasan\footnote{mzhasan@princeton.edu}}
\affiliation { Laboratory for Topological Quantum Matter and Spectroscopy (B7), Department of Physics,  Princeton University, Princeton, New Jersey 08544, USA}


\date{\today}

\begin{abstract}

Superconductivity in topological band structures is a platform for realizing Majorana bound states and other exotic physical phenomena such as emergent supersymmetry. This potential nourishes the search for topological materials with intrinsic superconducting instabilities, in which Cooper pairing is introduced to electrons with helical spin texture such as the Dirac surface states of topological insulators, forming a time-reversal symmetric topological superconductor on the surface. We employ first-principles calculations and ARPES experiments to reveal that PbTaSe$_2$, a non-centrosymmetric superconductor, possess a nonzero $\mathbb{Z}_2$ topological invariant and fully spin-polarized Dirac surface states. Moreover, we analyze the phonon spectrum of PbTaSe$_2$ to show how superconductivity can emerge due to a  stiffening of phonons by the Pb intercalation, which diminishes a competing charge-density-wave instability. Our work establishes PbTaSe$_2$ as a stoichiometric superconductor with nontrivial $\mathbb{Z}_2$ topological band structure, showing it holds great promise for studying aspects of topological superconductors such as Majorana zero modes.

\end{abstract}

\maketitle


The interplay of topology and interacting electronic instabilities promises new ways to realize exotic physical phenomena. Majorana zero modes have been theoretically proposed to exist in the vortex core of chiral $p$-wave superconductors \cite{TS-4,TS-1,TS-2,TS-3,TS-5,TS-6,TS-7,TS-8,TS-9,MF,Das-1,Das-2}. These systems are topological due to the unconventional form of their order parameter, while their electronic band structure can be topologically trivial. Unfortunately, electrons rarely form a chiral $p$-wave superconducting order in real materials. In a seminal work, Fu and Kane~\cite{TS-2} realized that this can be overcome by interchanging the role of band structure and order parameter: Majorana zero modes can also appear when the topological surface states of a $\mathbb{Z}_2$ topological band structure are gapped due to ordinary $s$-wave superconducting pairing. Multiple routes that have been proposed to realize this idea. One is to dope topological insulators (TIs), for example with 
copper atoms, giving rise to a transition temperature $T_{c}$ up to 3.8 K in  Bi$_2$Se$_3$ \cite{TS-dop1,TS-dop2,TS-dop3}. However, this approach requires fine tuning of the chemical doping composition and also introduces in an uncontrollable way chemical disorder into the system. Besides, the nature of the resulting superconductivity remains unclear. Another approach is to artificially fabricate TI thin layers on top of a superconductor \cite{TScon}. Helical pairing of Dirac surface states induced by the proximity effect has been observed experimentally \cite{TScon}. However, complex interface interaction hinders a comprehensive understanding of the induced helical Cooper pairing  and the superconducting gap drops steeply as the film thickness increases, which places stringent requirements on material synthesis.  A way that overcomes these difficulties is to search for chemically stoichiometric superconductors which possess intrinsically topological surface states. To date, a stoichiometric superconductors with nontrivial topological band structure remains elusive in real materials. 

In this paper, we use first-principles calculations and ARPES experiments to show that PbTaSe$_2$, a recently discovered non-centrosymmetric superconductor~\cite{PTS-exp}, possesses a nonzero $\mathbb{Z}_2$ topological invariant and topologically protected surface states. PbTaSe$_2$ is part of the family of layered transition metal dichalcogenide (TMD) superconductors, several of which show a competition between charge density wave (CDW) and superconducting order~\cite{Klemm15}. For example, in Cu$_x$TiSe$_2$ a superconducting dome appears to cover the quantum critical point at the Copper doping at which the CDW phase terminates~\cite{CDW-4}. To understand how superconductivity can compete with the CDW instability in PbTaSe$_2$, we performed a systematic phonon calculation, revealing a surprisingly strong electron-phonon coupling associated with the $A_{1g}$ phonon mode that accompanies the vertical motion of Pb atoms. From the theoretically obtained electron-phonon coupling strength, we estimate the 
superconducting transition temperature to be 3.1 K, which agrees very well with our experimental measurements on single crystal PbTaSe$_2$. Taken together the BCS-like superconductivity and Dirac surface states of PbTaSe2, our work established, for the first time, a material realization of a stoichiometric superconductor with topological band structure, providing a  promising platform for studying exotic Majorana physics in condensed matter.

\begin{figure}
\centering
\includegraphics[width=16cm]{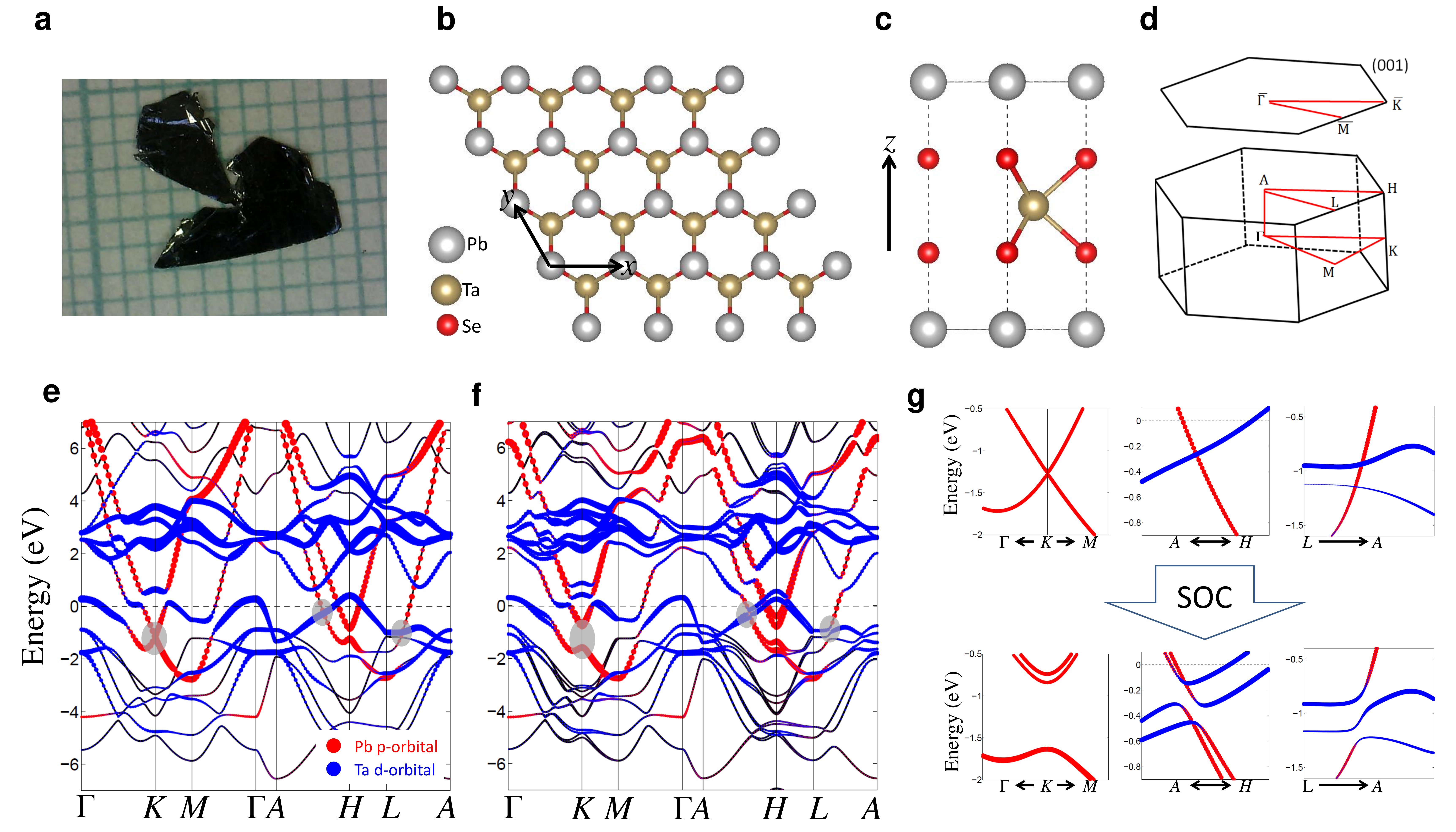}
\caption{
({a}) Optical image of PbTaSe$_2$ single crystals measured in this work. 
({b, c}) Top view and side view of PbTaSe$_2$ crystal structure. The silver, yellow, and red balls indicate Pb, Ta, and Se atoms, respectively.
({d}) Bulk and (001) surface Brillouin zone (BZ) of PbTaSe$_2$. 
({e}) Bulk band structure using the method of generalized gradient approximation (GGA). Red-dotted lines present electron-like Pb-$6p$ bands  band dispersion. Ta-$5d$ bands (blue-dotted lines) are hole-like near the Fermi level, which hybridize strongly with Pb $6p$-bands. 
({f}) Band structure calculated by GGA+SOC. SOC opens a continuous energy gap in the whole BZ highlighted by the gray areas. 
({g}) Zoom-in band structures of the gray areas of GGA [panel ({d})] and GGA+SOC [panel ({e})]. The band crossing is gapped out by SOC, forming a continuous energy gap across the entire BZ.
}
\label{band}
\end{figure}

\begin{figure}
\centering
\includegraphics[width=16cm]{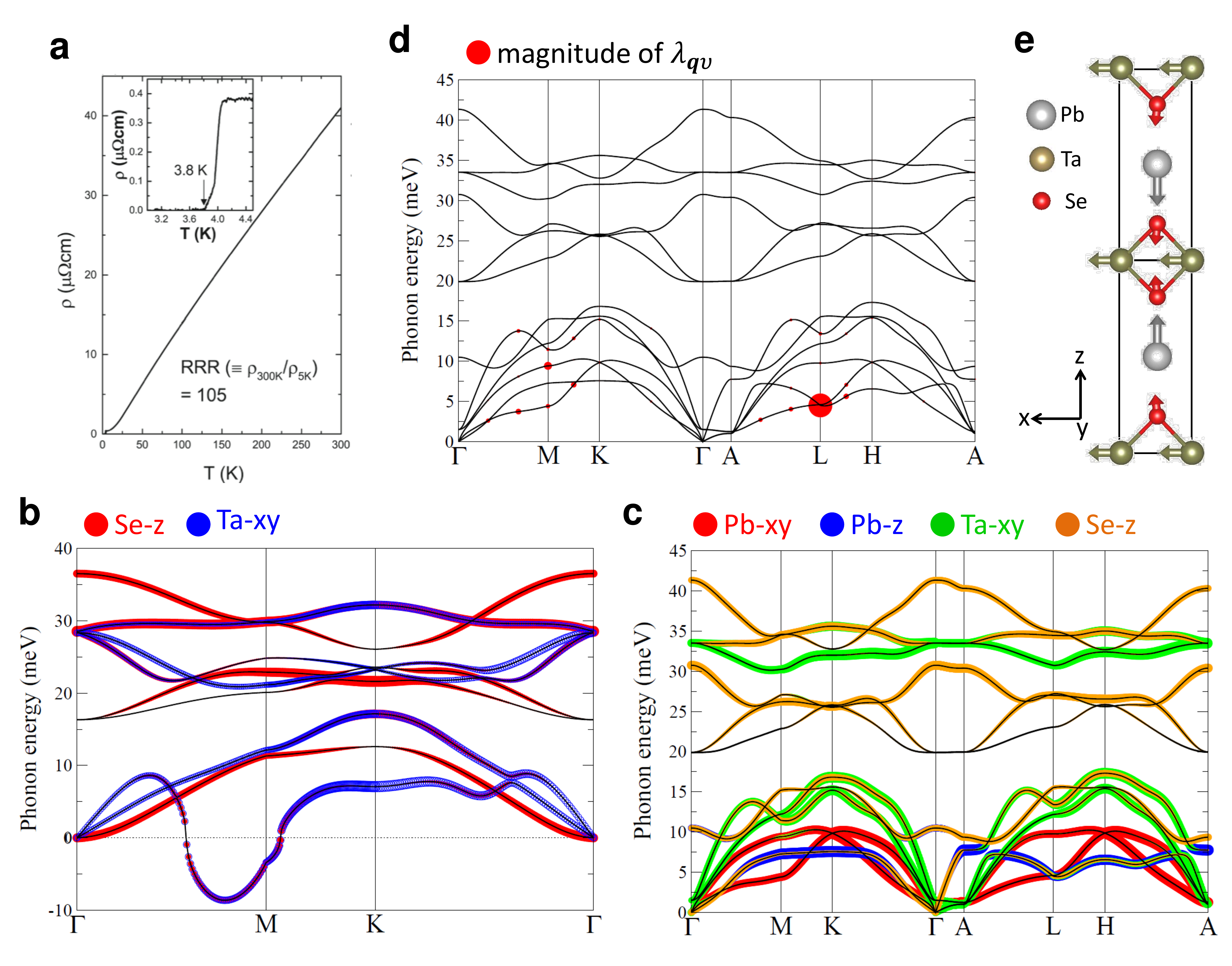}
\caption{
({a}) Measured resistivity as a function of temperature showing a superconducting transition temperature $T_c$ = 3.8 K. 
({b}) Phonon spectrum of monolayer TaSe$_2$. The negative-energy part, which is related to phonons that involve the vertical (horizontal) motions of Se (Ta), indicates the CDW instability of monolayer TaSe$_2$. The red and blue dots indicate Se vertical mode and Ta horizontal mode, respectively. 
({c}) Phonon spectrum of PbTaSe$_2$ weighted by the motion modes of Pb, Ta and Se atoms. The red, blue, green and orange dots indicate Pb horizontal, Pb vertical, Ta horizontal and Se vertical modes, respectively. 
({d}) Phonon spectrum of PbTaSe$_2$ weighted by the magnitude of the electron-phonon coupling $\lambda_{{\bs q}\nu}$ (the size of red dots). 
({e}) Atomic vibration pattern of the phonon mode at $L$ with surprisingly strong electron-phonon coupling. The out-of-phase atomic displacements in z direction are explicitly shown in the 1$\times$1$\times$2 supercell.
}
\label{ep}
\end{figure}

PbTaSe$_2$ crystallizes in a structure of alternately-stacked hexagonal Pb and TaSe$_2$ layers with space group P$\bar{6}$m2 [Fig. \ref{band}(b)]. In a TaSe$_2$ layer, a hexagonal Ta atomic plane is sandwiched by two hexagonal Se atomic planes, forming strong ionic bonds within a local trigonal structure. The lattice can also be viewed as a Pb layer intercalating two adjacent TaSe$_2$ layers with Pb atoms sitting right above Se atoms. The Pb-Se bonds are relatively weak compared to the ionic Ta-Se bonds. This soft bonding plays an important role in the emergent superconductivity in PbTaSe$_2$, as will be discussed later. For this study, PbTaSe$_2$ single crystals were prepared by the chemical vapor transport (CVT) method, as shown in Fig. \ref{band}(a). 
We performed a transport measurement to verify the superconducting property of our samples. Figure \ref{ep}(a) shows the resistivity curve displaying a clear superconducting transition at a temperature of 3.8~K, which is consistent with previous work on polycrystalline samples \cite{PTS-exp}. Before discussing the superconductivity, we will present the bulk band structure obtained from first-principles calculations, which help us to understand the electronic structure of PbTaSe$_2$. 

Figure \ref{band}(d) shows the bulk band structure of PbTaSe$_2$, calculated with the GGA method without the inclusion of spin-orbit coupling (SOC). The Pb $6p$-orbitals and Ta $5d$-orbitals dominate the  bands around the Fermi energy $E_\mathrm{F}$, while Se 2$p$ states are mainly below $-2$~eV. The Pb $6p$-bands display electron-like band dispersion crossing $E_\mathrm{F}$ with a bandwidth larger than 7~eV and form a Dirac bulk band dispersion at the $K$ point below the Fermi level [marked by gray areas in Fig. \ref{band}(d)]. In contrast to Pb $6p$-bands, Ta $5d$-orbitals exhibit a quite different behavior. The bands from Ta $5d$-orbitals split into two groups. The upper part lies at +3~eV above $E_\mathrm{F}$ with a narrow bandwidth about 2~eV. The lower part displays hole-like band dispersion near $E_F$ and hybridizes strongly with Pb $6p$-bands around $K$ and $H$ points, leading to a complex band inversion feature around E$_\mathrm{F}$. Note that the band dispersion along $\Gamma$--$K$--$M$ ($k_z$ = 0)
 is very different from that along $A$--$H$--$L$ ($k_z$ = $\pi$), which indicates the coupling along $z$ direction can not be neglected even though the crystal is comprised of two-dimensional layers \cite{PTS-exp}. This non-negligible interlayer coupling along the $z$ direction results in a surprising large electron-phonon coupling at $L$ , which will be discussed later. When SOC is turned on, the band crossings in Fig. \ref{band}(d) are gapped out [shown in the gray areas in Fig. \ref{band}(e) and the zoom-in shown in Fig.1 (F)] resulting in a continuous energy band gap spreading over the whole Brillouin zone (BZ). The band structure shows significant spin-orbit splitting due to the breaking of spatial inversion symmetry, except at the time-reversal invariant points ($\Gamma$, $M$, $A$, and $L$) of the BZ. 


To understand the emergence of superconductivity in PbTaSe$_2$, we performed systematic calculations for the electron-phonon couplings. Since bulk 2$H$-TaSe$_2$ is known to exhibit a CDW instability below 122K \cite{CDW-1,CDW-2}, we first examine the stability of monolayer (ML) TaSe$_2$ which is a building block of PbTaSe$_2$. As shown in Fig.~\ref{ep}(b), the CDW instability is also seen in ML TaSe$_2$. This instability results from the negative phonon energy of the $A_{1g}$ mode that involves the horizontal motions of Ta and vertical motions of Se. Figure~\ref{ep}(c) shows the phonon spectrum of PbTaSe$_2$. The CDW instability is eliminated by the intercalation of Pb layer.  The phonon spectrum in Fig.~\ref{ep}(c) is displayed with the decomposition of atomic vibrations. The soft phonons stem mostly from the vibration of Pb atoms. The energy of the TaSe$_2$-related phonons are pushed upward, partly due to the fact that the originally unstable phonons that involve the vertical motions of Se are stiffened by the Pb intercalation. We also note that the energy of phonons along $\Gamma$--$A$ is very low, which implies that the Pb intercalation moves the system to the vicinity of another CDW instability that corresponds to the relative sliding between the TaSe$_2$ slabs and the Pb sheets. This can be understood from the fact that the vertical bonding between TaSe$_2$ and Pb-$p_x$/$p_y$ orbitals is very weak. In Fig.~\ref{ep}(d), the strength of electron-phonon coupling is visualized by red dots to reveal the $\boldsymbol{q}$-dependence of the electron-phonon coupling ($\lambda_{\boldsymbol{q}\nu}$). In our calculations, we find the total electron-phonon coupling $\lambda =  \sum_{\boldsymbol{q}\nu}\lambda_{\boldsymbol{q}\nu}$ to be $\lambda=0.66$. Using the the McMillan formula (please see Method section for details), the superconducting transition temperature $T_c$ is estimated to be 3.1 K. Our calculation agrees well with the results from the polycrystalline transport and specific 
heat experiments \cite{PTS-exp} ($\lambda$ = 0.74 and $T_{c} \sim$ 3.8 K) and our single crystal transport measurement, indicative of a BCS-like superconductivity in PbTaSe$_2$. As shown in Fig.~\ref{ep}(d), the major contribution to the effective electron-phonon coupling comes from the $A_{1g}$ phonons at $L$. This wave vector does not lead to any apparent nesting of the band structure, which might be a reason why it does not trigger a CDW instability. It is also worth noting that this surprisingly large coupling strength takes place at $L$, but not at $M$. The reason for this difference lies in their influence on the Se-Ta-Se bond angle when the $A_{1g}$ phonons vibrate with different wavevector $q_z$. At $L$, the two Pb atoms that sandwich the TaSe$_2$ slab move in opposite directions [Fig. \ref{ep}(e)], which greatly alters the Se-Ta-Se bond angle. At $M$, on the other hand, the in-phase motions of these two Pb atoms in $z$ direction exert less influence on the Se-Ta-Se bond angle. In general, the bond angle has been shown to yield modifications of the electronic structures of TMDs \cite{TMD-2}. The enhanced change of the bond angle induced by the $A_{1g}$ phonon at $L$ explains the large effective electron-phonon coupling. 

From our analysis of the phonon spectrum, we conclude that the emergence of superconductivity in PbTaSe$_2$ can be attributed to the disappearance of the CDW. This is a common physical mechanism in TMDs where the CDW and superconductivity compete with each other \cite{CDW-3,CDW-4,CDW-5,CDW-6,CDW-7}. Mostly, the CDW predominates in the competition for the gapping of Fermi surfaces. Superconductivity takes place when the CDW is suppressed by either pressurization or doping \cite{CDW-4,CDW-5,CDW-6,CDW-7}. In PbTaSe$_2$, the phonon stiffening due to the Pb intercalation makes it energetically unfavorable to form the CDW order, opening the way for the transition into a superconducting phase.

We now turn to the topological aspects of the electronic structure of PbTaSe$_2$. In most typical three-dimensional TIs (such as Bi$_2$Se$_3$) the non-trivial topological bands originate from a band inversion induced by the SOC. The scenario in PbTaSe$_2$ is rather different: an inverted band structure with two different orbital characters (Pb-$p$ and Ta-$d$) exists in PbTaSe$_2$ even without SOC as the highly dispersive Pb and Ta bands cross each other close to the Fermi level. The role of SOC in PbTaSe$_2$ is to gap out the Dirac-like band dispersion at $K$ and other band crossings in the BZ [highlighted by gray shades in Fig. \ref{band}(f)], forming a continuous energy gap across the whole BZ. This is analogous to graphene in which the Dirac band at the $K$ point is gapped by the intrinsic SOC resulting in a quantum spin Hall phase, while the band structure is inverted at the $M$ points already in absence of SOC \cite{TI-G}. Hence, even though PbTaSe$_2$ has a metallic ground state, in presence of SOC the 
valence band remains separated from the conduction band by a continuous gap throughout the whole BZ. This allows to associate a well-defined $\mathbb{Z}_2$ topological invariant with the band structure~\cite{TI-G}. Since PbTaSe$_2$ lacks inversion symmetry, the $\mathbb{Z}_2$ topological invariant cannot be computed via a parity analysis \cite{Z2-parity}. For this reason, we calculated the $\mathbb{Z}_2$ index by the hybrid Wannier functions method \cite{Z2method}. The $\mathbb{Z}_2$ topological number is extracted from a first-principles calculation using the Quantum Espresso code \cite{QE} within the framework of the local density approximation (LDA) with SOC taken into account. A 30$\times$30$\times$15 $\boldsymbol{k}$-grid is used to sample the BZ. The calculation shows a non-zero $\mathbb{Z}_2$ number ($\mathbb{Z}_2 = 1$), indicating that PbTaSe$_2$ is in the topological insulator phase.

\begin{figure}
\centering
\includegraphics[width=16cm]{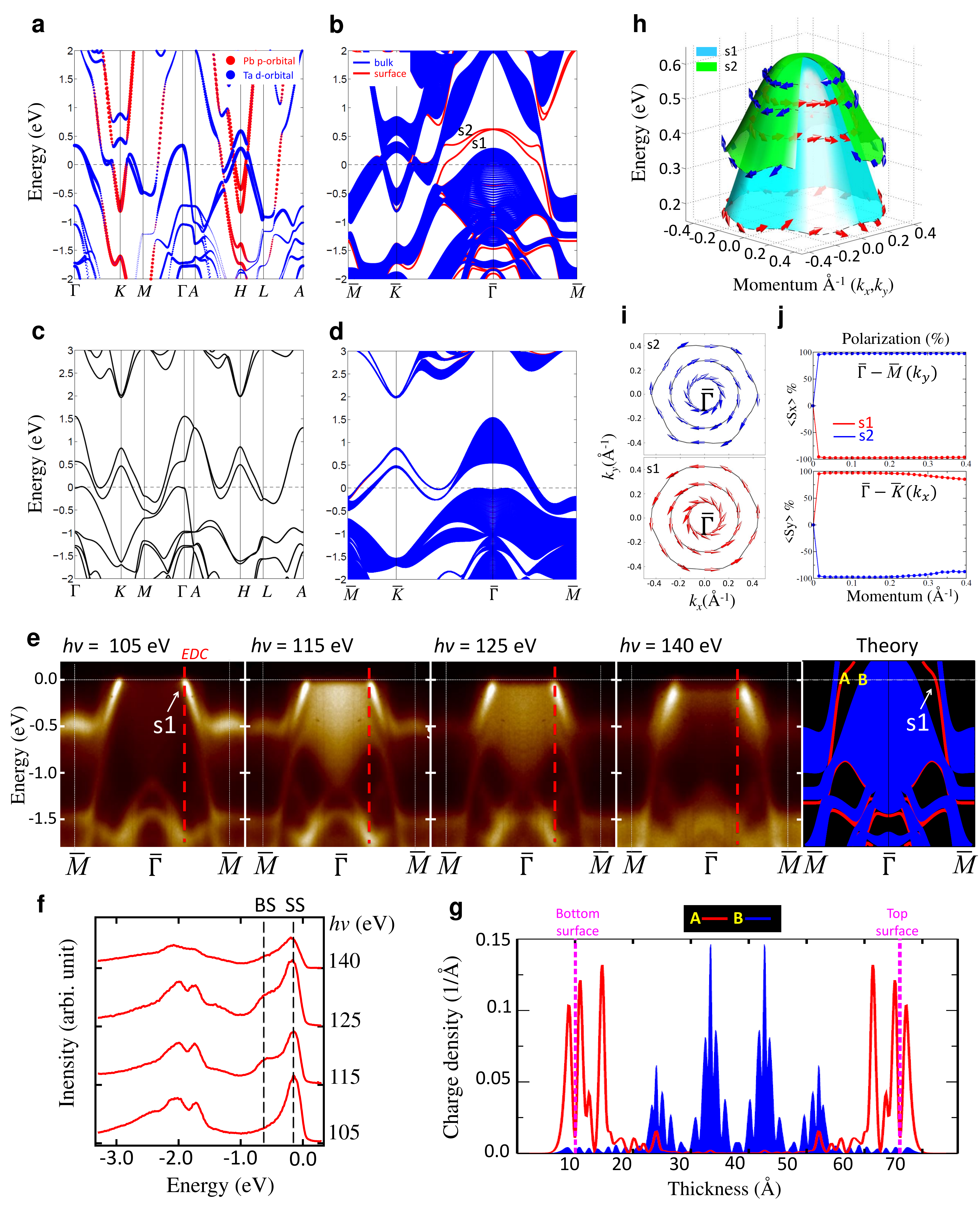}
\caption{
({a}) Bulk band structure of PbTaSe$_2$ by GGA+SOC. 
({b}) Band structure of a 6-unit-cell PbTaSe$_2$ symmetric slab. The surface states ($s1$ and $s2$) and bulk band projections onto (001) surface are represented by the red and blue lines, respectively. }
\label{ss}
\end{figure}

Non-trivial topology of bulk band structure guarantees the existence of gapless surface states which are the hallmark of topological insulators. However, it is not clear a priori whether these surface states are observable if the band structure is metallic. The projection of the bulk bands in the surface BZ could cover the topological surface states. In the following we present first principle calculations and ARPES measurements showing that the surface states can indeed be observed in the (001) surface of PbTaSe$_2$ thanks to the layered nature of the material. Figure~\ref{ss}(b) shows the surface band structure of PbTaSe$_2$ calculated using a 6-unit-cell symmetric slab. Red and blue lines represent surface and bulk states projected onto the (001) surface [Fig. \ref{band}(c)], respectively. The gapless surface band lies in the continuous energy gap around the $\bar{\Gamma}$ point. Along $\bar{\Gamma}$--$\bar{M}$ and $\bar{\Gamma}$--$\bar{K}$, theres is only one surface band (red line) connecting the 
valence and conduction bands (blue lines), which is the signature of the nontrivial topological bands. Figure~\ref{ss}(e) shows ARPES spectra of PbTaSe$_2$ taken with different photon energies along $\bar{\Gamma}$--$\bar{M}$ direction. Near $\bar{\Gamma}$ the intensity from the hole-like pocket varies prominently with photon energy, clearly indicating that it is part of the bulk band structure. In contrast, the band labelled s1 poking the Fermi level in between $\bar{\Gamma}$ and $\bar{M}$ shows a dispersion independent of photon energy, which evidences the surface nature of this band. We plot energy distribution curves [EDC, Fig.~\ref{ss}(f)] along the lines marked in Fig.~\ref{ss}(e). At different energies, the position of peak SS is unchanged, further proving the surface feature of band s1. By contrary the peak BS shows up at 115 eV but disappears at 105 and 140 eV, reflecting the bulk nature of this state. The ARPES results are remarkably consistent with theoretical prediction. Therefore, our ARPES measurements provide strong evidence for the existence of topological surface states in PbTaSe$_2$. The calculated charge density distribution (in real space) of states A and B [marked in Fig.~\ref{ss}(e)] is shown in Fig.~\ref{ss}(g). Apparently, state A in band s1 is a typical surface state while state B is a bulk state.

A spin-momentum-locked helical spin-texture is one of most prominent characteristics of topological surface states. The cyan and green bell-like contours in Fig.~\ref{ss}(h) show a three-dimensional sketch of the surface states $s1$ and $s2$ around the $\bar{\Gamma}$ point, respectively. The spin-textures can be seen clearly in the two-dimensional energy contours [Fig.~\ref{ss}(i)]. The in-plane spin-texture surrounding $\bar{\Gamma}$ displays the right- and left-handed chirality for $s1$ and $s2$, respectively, which is same as the surface spin-texture of Bi$_2$Se$_3$ \cite{TI-spin}. The spin polarizations at $+\bs{k}$ and $-\bs{k}$ are opposite to each other, as dictated by time-reversal symmetry. Away from the $\bar{\Gamma}$ point, the surface bands show hexagonal warping~\cite{warping}. This effect also leads to the finite out-of-plane spin component $\langle S_{z}\rangle$ along the $\bar{\Gamma}$--$\bar{K}$ ($k_x$) direction [Fig.~\ref{ss}(h)]. In contrast, the $\langle S_{z}\rangle$ must be zero along 
$\bar{\Gamma}$--$\bar{M}$ ($k_y$) because of the mirror symmetry of the crystal structure~\cite{warping}. We note that the spin polarization of surface states of PbTaSe$_2$ around $\bar{\Gamma}$ point is nearly 100$\%$ [Fig.~\ref{ss}(J)]. The in-plane spin polarization in PbTaSe$_2$ remains nearly 100$\%$ along $\bar{\Gamma}$--$\bar{M}$ even for a large momentum $k_{x}$ $\sim$ 0.3 \AA$^{-1}$, but decreases slightly along $\bar{\Gamma}$--$\bar{K}$ ($k_y$) direction due to the emergence of the finite $\langle S_{z}\rangle$ polarization from the hexagonal warping effect \cite{warping}.

To highlight the critically important role of the Pb 6$p$-orbitals in the formation of the non-trivial topological phase in PbTaSe$_2$, we calculated the electronic structure of TaSe$_2$ for comparison. Figure~\ref{ss}(c) shows bulk band structure of TaSe$_2$. Contrary to PbTaSe$_2$, which shows complex band inversion around $E_\mathrm{F}$ [Fig.~\ref{ss}(a)], TaSe$_2$ displays a clear separation between the valence band and the conduction band.  Close to the $E_\mathrm{F}$ there exist two large hole pockets, one at $\Gamma$ and the other one at $K$. Those states are mainly derived from the Ta $5d$-orbitals. The band dispersion of TaSe$_2$ is similar to the Ta part in the bulk band of PbTaSe$_2$ [blue dots in Fig.~\ref{ss}(a)]. We also checked that the $\mathbb{Z}_2$ invariant in TaSe$_2$ is trivial ($\mathbb{Z}_2=0$). To further confirm this result, we performed slab calculation for TaSe$_2$ (Fig.3(d)). It shows a clear energy gap with no surface band connecting the valence band and the conduction 
band, consistent the trivial $\mathbb{Z}_2$ index.

Let us finally discuss what promises the topological surface states of PbTaSe$_2$ hold for observing Majorana zero modes in vortices.
 The two crucial ingredients needed are helical surface states that are well separated in momentum space from all bulk states and a fully gapped superconducting order. We have demonstrated that PbTaSe$_2$ possesses the first ingredient. Figures~\ref{FS}(a), (b) show the calculated (001) Fermi surface and the ARPES measurement with 64~eV photon energy, respectively, clearly exposing the topological surface state. At Fermi level, there exist bulk band pockets surrounding at $\bar{\Gamma}$ and $\bar{K}$, but the topological surface states encircling $\bar{\Gamma}$ are not degenerate with these bulk bands, showing an isolated surface band with helical spin texture. As for the second ingredient, PbTaSe$_2$ is intrinsically superconducting without the aid of doping or artificial heterostructures. Our phonon calculation illustrates the superconductivity of PbTaSe$_2$ arises from an enormously enhanced electron-phonon coupling at $L$ and the estimated transition temperature is in accordance with the previous transport and specific heat measurements, indicating BCS-like Cooper pairing exists in this compound. When the temperature is below $T_c$, the topological surface states can be superconducting as a consequence of surface-bulk proximity effect as schematically shown in Figs.~\ref{FS}(c-e).  It is known that a superconducting spin-momentum-locked two-dimensional electron gas is topologically nontrivial with a helical (nodeless) Cooper pairing \cite{TScon}. Recently it has been confirmed by the thermal conductivity and specific heat measurements \cite{R1,R2} that the superconducting gap of PbTaSe$_2$ is nodeless, consistent with the picture of coexistence of BCS-type bulk superconducting states and helical surface superconducting states in PbTaSe$_2$. Furthermore, the proximity-induced surface superconductivity in a bulk compound such as PbTaSe$_2$ can be stronger than in an artificial TI-SC heterostructure since in an intrinsic superconductor there exists neither attenuation due to the finite thickness of the TI layer nor the complex interfacial condition \cite{TS-2,TScon}. To show the superconducting proximity effect, we performed a simulation with an effective Hamiltonian for a 12-layer PbTaSe$_2$ slab. In the model an on-site intra-orbital mean-field superconducting order parameter of $s$-wave spin singlet type was introduced. In order to clearly demonstrate the proximity effect on the surface states the bare pairing potential is set to decay to zero in the surface layer [Fig.~\ref{FS}(f)]. Due to the extend of the surface states into the bulk, a superconducting gap is induced in surface bands. In our model we chose the bulk gap to be 50 meV, resulting in a proximitized nodeless gap of 5 meV on the surface states [Fig.~\ref{FS}(g)]. Thanks to this hybridization with bulk states and spin-momentum locking of the surface states, helical topological superconductivity can be induced on the surface \cite{TS-2,TScon}. Therefore, PbTaSe$_2$, an intrinsic superconductor with topological surface states, is a potential platform to study Majorana modes in the core of vortices \cite{PRL1, PRL2, PRB}.

In summary, we have investigated electronic structure and electron-phonon coupling of PbTaSe$_2$ via first-principles calculations and ARPES experiments. The CDW existing in TaSe$_2$ is eliminated in PbTaSe$_2$ by the intercalation of Pb layers. We also uncovered that the electron-phonon coupling is dominated by the $A_{1g}$ mode at the $L$ point that accompanies the squeezing of the TaSe$_2$ layer. The theoretical electron-phonon coupling strength $\lambda$ is about 0.66 and estimates $T_c$ to be 3.1~K, consistent with our single crystal transport measurement ($T_c \sim$ 3.8 K). For the electronic structure, the Pb-$6p$ orbitals hybridize strongly with Ta-$5d$ orbitals at the $K$ and $H$ points, leading to a well-defined band-inverted energy gap and, consequently, a topologically nontrivial  band structure. Slab calculations evidence the topological Dirac surface states with spin-momentum-locked spin texture. The calculated dispersion of the Dirac surface states is in good agreement with our ARPES measurement. Our work establishes that PbTaSe$_2$, a stoichiometric compound, is a superconductor with nontrivial band topology, potentially providing a platform for studying Majorana bound states in vortices.

$Note:$ A very recent experimental work \cite{nodeless} has shown that PbTaSe$_2$ has a nodeless superconducting gap which is consistent with our theoretical discussion reported here.

\section{Methods}
We computed electronic structures using the projector augmented wave method \cite{PAW-1,PAW-2} as implemented in the VASP \cite{VASP-1,VASP-2,VASP-3} package within the generalized gradient approximation (GGA) \cite{PBE} schemes. Experimental lattice constants were used \cite{expstr}. A 12$\times$12$\times$4 MonkhorstPack $k$-point mesh was used in the computations. The SOC effects are included self-consistently. 
The electron-phonon coupling $\lambda_{{\bs q}\nu}$ is computed 
based on density functional perturbation theory \cite{DFPT}
implemented in the Quantum Espresso code \cite{QE},
using 
\begin{equation}
\lambda_{{\bs q}\nu} = \frac{1}{\pi N_F}\frac{\Pi^{''}_{{\bs q}\nu}}{\omega_{{\bs q}\nu}^{2}},
\end{equation}
where $N_F$ is the density of states (DOS) at the Fermi level, and
$\omega_{{\bs q}\nu}$ is the phonon frequency of mode $\nu$
at wave vector $\bs q$.
The electron-phonon quasiparticle linewidth, $\Pi^{''}_{{\bs q}\nu}$, is given by
\begin{equation}
\Pi^{''}_{{\bs q}\nu} = \pi\omega_{{\bs q}\nu}\sum_{mn,{\bs k}}|\textbf{g}^{\nu}_{mn}({\bs k},{\bs q})|^2\delta(\epsilon_{n{\bs k}})\delta(\epsilon_{m{\bs k}+{\bs q}}),
\end{equation}
where $\epsilon_{n{\bs k}}$ is the energy of the KS orbital and the dynamical matrix reads
\begin{equation}
\textbf{g}^{\nu}_{mn}({\bs k},{\bs q}) = \left(\frac{\hbar}{2\mathrm{M}\omega_{{\bs q}\nu}}\right)^{\frac{1}{2}}\langle\psi_{n{\bs k}}|\frac{dV_{scf}}{d\textbf{u}_{{\bs q}\nu}}\cdot \hat{\textbf{e}}_{{\bs q}\nu}|\psi_{m{\bs k}+{\bs q}}\rangle,
\end{equation}
where $\frac{dV_{scf}}{d\textbf{u}_{{\bs q}\nu}}$ 
represents the deformation potential at the small atomic displacement
$d\textbf{u}_{{\bs q}\nu}$ of the given phonon mode. 
M and $\hat{\textbf{e}}_{{\bs q}\nu}$ denote the mass of the atom 
and the unit vector along $\textbf{u}_{{\bs q}\nu}$, respectively. 
The critical temperature $T_c$ can then be estimated by the McMillan formula:
\begin{equation}
T_c = \frac{\mathbf{\omega}_{ln}}{1.20}\exp{[-\frac{1.04(1+\lambda)}{\lambda-\mu^{*}(1+0.62\lambda)}]},
\end{equation}
where
\begin{eqnarray}
\lambda & = & \sum_{{\bs q}\nu}\lambda_{{\bs q}\nu}, \\
\omega_{ln} & = & \exp{[\frac{2}{\lambda}\int{d\omega\frac{ln(\omega)}{\omega}}\alpha^{2}F(\omega)]}, \\
\alpha^{2}F(\omega)& = & \frac{1}{2}\int_{BZ}{d\omega\lambda_{{\bs q}\nu}\omega_{{\bs q}\nu}\delta(\omega-\omega_{{\bs q}\nu})},
\end{eqnarray}
and $\mu^*$ is chosen to be 0.1 eV in the estimation of $T_c$.

We simulated the superconducting state in a 12-layered PbTaSe$_2$ slab. The non-superconducting slab Hamiltonian is constructed from the Wannier-function-based tight-binding model and the superconducting order parameter is chosen to be on-site intra-orbital spin singlet pairing among Pb $p$ and Ta $d$-orbitals. For simplicity, we assumed equal gap amplitude for all orbitals. The boundary effect was considered by assuming the gap function in the $n_z$-th layer to be 
\begin{eqnarray}
\Delta(n_z) = \Delta_0 \left(1 - \frac{2}{\pi} \arctan{\left(\frac{\xi}{n_z-1} + \frac{\xi}{N_z-n_z} \right)} \right)
\end{eqnarray}
($n_z=1, \,2, \, ..., \, N_z=12$). $\Delta_0$ is chosen such that $\Delta$=50 meV at the center of the slab and $\xi=1$ is used. The choice of the gap function (8) produces a zero pairing at the boundaries ($n_z=1$ and $n_z=12$).

 ARPES measurements were performed at the liquid nitrogen temperature in the beamline I4 at the MAX-lab in Lund, Sweden.  The energy and momentum resolution was better than 20 meV and 1$\%$ of the surface Brillouin zone (BZ) for  ARPES measurements at the beamline I4 at the MAX-lab. Samples were cleaved in situ under a vacuum condition better than 1 $\times$ 10$^{-10}$ torr. Samples were found to be stable and without degradation for a typical measurement period of 24 hours. 
 
 Single crystals of PbTaSe$_2$ were grown by the CVT method using chlorine in the form of PbCl$_2$ as a transport agent. For the pure synthesis of PbTaSe$_2$, stoichiometric amounts of the elements (purity of Pb and Ta: 6N, of Se: 5N) were loaded into a quartz ampoule, which was evacuated, sealed and fed into a furnace (850$\celsius$) for 5 days. About 10 g of the prereacted  PbTaSe$_2$ were placed together with a variable amount of PbCl$_2$ (purity 5N) at one end of another silica ampoule (length 30-35 cm, inner diameter 2.0 cm, outer diameter 2.5 mm). All treatments were carried out in an Argon box, with continuous purification of the Argon atmosphere resulting in an oxygen and water content of less than 1 ppm. Again, the ampoule was evacuated, sealed and fed into a furnace. The end of the ampoule containing the prereacted material was held at 850$\celsius$  , while the crystals grew at the other end of the ampoule at a temperature of 800$\celsius$ (corresponding to a temperature gradient of 2.5 K/cm) during a time of typically 1 week. Compact single crystals of sizes of up to 8 $\times$ 5 $\times$ 5 mm$^3$ were obtained.

\section{Acknowledgements}
T.R.C. acknowledges visiting scientist support from Princeton University. We gratefully acknowledge C. M. Polley, J. Adell, M. Leandersson, T. Balasubramanian for their beamline assistance at the Maxlab. We also thank Chen Fang, Ching-Kai Chiu and Andreas P. Schnyder for discussions.

\newpage

\newpage

\newpage

\addtocounter{figure}{-1}
\begin{figure*}[t!]
\caption{(c) Bulk band structure of 2$H$-TaSe$_2$ by GGA+SOC. The Ta-5$d$ orbitals dominated in the bands around the E$_F$. 
({d}) Band structure of a 12-unit-cell TaSe$_2$ symmetric slab. 
({e}) ARPES mappings taken with photon energy from 105 to 140 eV and theoretical band structure of PbTaSe$_2$ along the $\bar{\Gamma}$-$\bar{M}$ direction.
({f}) Energy distribution curves (EDC) along red dash lines marked in (E).
({g}) Real space charge density distribution of state A and B as marked in (E).
({h})  Calculated energy surface and spin-texture (red and blue arrows) of surface states ($s1$ and $s2$) of PbTaSe$_2$. 
({i})~Top view of spin-texture of surface states with different binding energies. 
({j}) Spin polarization of surface states along $\bar{\Gamma}$-$\bar{M}$ and $\bar{\Gamma}$-$\bar{K}$, indicating nearly full spin polarization close to $\bar{\Gamma}$ point.}
\end{figure*}

\clearpage
\newpage

\begin{figure}
\centering
\includegraphics[width=16cm]{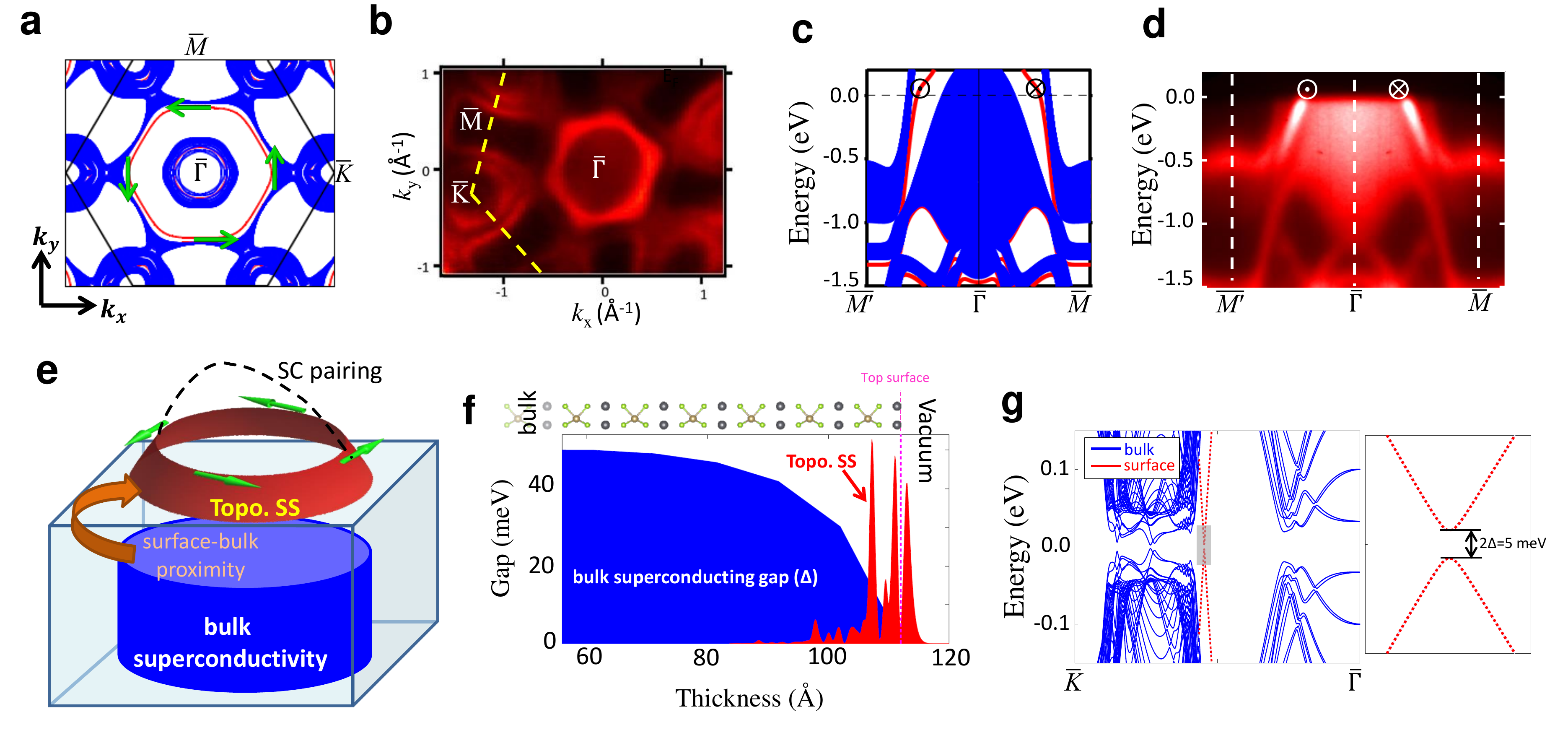}
\caption{
({a}) DFT Fermi surface contour of PbTaSe$_2$ (001) surface and 
({b}) ARPES Fermi surface taken with 64 eV photons. 
({c}) Band structure cut along ${\bar{M'}}$-$\bar{\Gamma}$-${\bar{M}}$. The spin orientation of the surface states is indicated.
({d}) APRES spectral cut along ${\bar{M'}}$-$\bar{\Gamma}$-${\bar{M}}$. 
({e}) Schematic of helical Cooper pairing as a consequence of the surface-bulk proximity effect. 
({f}) Profile of the superconducting gap function (blue area) and topological surface states (red area) in a 12-layer PbTaSe$_2$ slab model.
({g}) Band structure of a 12-layer PbTaSe$_2$ slab model with the on-site superconducting gap in panel ({\bf f}). The (001)-surface and bulk bands are shown by the red and blue lines, respectively. The zoom-in shaded area is shown in the right side panel.
}
\label{FS}
\end{figure}

\end{document}